\documentclass[reprint, secnumarabic, amssymb, amsmath, superscriptaddress, aps, prl]{revtex4-1}
\usepackage{graphicx}
\usepackage[colorlinks=true,urlcolor=blue]{hyperref}
\usepackage{hyperref,ulem}
\usepackage{xcolor}
\hypersetup{
colorlinks,
linkcolor={red!50!black},
citecolor={blue!50!black},
urlcolor={blue!80!black}
}
\begin{document}

\title{Multiple charge density waves and superconductivity nucleation at antiphase domain walls in the nematic pnictide Ba$_{1-x}$Sr$_{x}$Ni$_{2}$As$_{2}$
}

\author{Sangjun Lee}
\affiliation{Department of Physics and Materials Research Laboratory, University of Illinois, Urbana, Illinois 61801, USA}
\author{John Collini}
\affiliation{Maryland Quantum Materials Center, Department of Physics, University of Maryland, College Park, Maryland 20742, USA}
\author{Stella X.-L. Sun}
\author{Matteo Mitrano}
\author{Xuefei Guo}
\affiliation{Department of Physics and Materials Research Laboratory, University of Illinois, Urbana, Illinois 61801, USA}
\author{Chris Eckberg}
\affiliation{Maryland Quantum Materials Center, Department of Physics, University of Maryland, College Park, Maryland 20742, USA}
\author{Johnpierre Paglione}
\affiliation{Maryland Quantum Materials Center, Department of Physics, University of Maryland, College Park, Maryland 20742, USA}
\affiliation{Canadian Institute for Advanced Research, Toronto, ON M5G 1Z8, Canada}
\author{Eduardo Fradkin}
\affiliation{Department of Physics and Materials Research Laboratory, University of Illinois, Urbana, Illinois 61801, USA}
\affiliation{Institute of Condensed Matter Theory, University of Illinois, Urbana, Illinois 61801, USA}
\author{Peter Abbamonte}
\email{abbamonte@mrl.illinois.edu}
\affiliation{Department of Physics and Materials Research Laboratory, University of Illinois, Urbana, Illinois 61801, USA}

\begin{abstract}
How superconductivity interacts with charge or nematic order is one of the great unresolved issues at the center of research in quantum materials. Ba$_{1-x}$Sr$_{x}$Ni$_{2}$As$_{2}$ (BSNA) is a charge ordered pnictide superconductor recently shown to exhibit a six-fold enhancement of superconductivity due to nematic fluctuations near a quantum phase transition (at $x_c=0.7$) \cite{Eckberg2020}. 
The superconductivity is, however, anomalous, with the resistive transition for $0.4 < x< x_c$ occurring at a higher temperature than the specific heat anomaly. Using x-ray scattering, we discovered a new charge density wave (CDW) in BSNA in this composition range. The CDW is commensurate with a period of two lattice parameters, and is distinct from the two CDWs previously reported in this material \cite{Lee2019,Eckberg2020}. We argue that the anomalous transport behavior arises from heterogeneous superconductivity nucleating at antiphase domain walls in this CDW. We also present new data on the incommensurate CDW, previously identified as being unidirectional \cite{Lee2019}, showing that is a rotationally symmetric, ``4$Q$" state with $C_4$ symmetry. Our study establishes BSNA as a rare material containing three distinct CDWs, and an exciting testbed for studying coupling between CDW, nematic, and SC orders.

\end{abstract}

\maketitle

One of the perennial questions in quantum materials is to what extent superconductivity (SC) may be enhanced by coupling to other Fermi surface instabilities, such as charge density wave (CDW), spin density wave (SDW) or nematic orders. This issue has been investigated most widely in cuprate and iron-based superconductors \cite{Fradkin2015,Tranquada2015, Paglione2010, Dai2015}. While spin fluctuations are considered a primary ingredient in Cooper pairing in both materials, the CDW in cuprates \cite{Abbamonte2005, Ghiringhelli2012, CominReview2016} and nematic fluctuations in iron-based materials \cite{Chu2012, Kuo2016} are pervasive and suggest a close interrelation between SC and these other electronic instabilities. The importance of such orderings for SC is still unclear so there is a great need for new materials that can provide fresh insight.

Ba$_{1-x}$Sr$_{x}$Ni$_{2}$As$_{2}$ is a nickel-based pnictide superconductor that is an ideal system to investigate the relationship between charge instabilities and SC. The parent compound, BaNi$_{2}$As$_{2}$, is a structural homologue of the prototypical iron-based superconductor BaFe$_{2}$As$_{2}$ with all Fe atoms replaced by Ni, that undergoes a first-order structural phase transition at $T_{\mathrm{s}} = 136$ K from tetragonal to triclinic symmetry \cite{Sefat2009, Eckberg2018}. We recently showed that BaNi$_{2}$As$_{2}$ is an intriguing material in which two distinct CDW instabilities arise sequentially on lowering temperature \cite{Lee2019}. In the tetragonal phase, an incommensurate CDW (IC-CDW), previously identified as unidirectional, forms along the in-plane $H$ axis with wave vector $q=0.28$ \cite{Lee2019}. Upon further cooling across the structural transition at $T_{\mathrm{s}}$, the IC-CDW is replaced by the second CDW, which we denote C-CDW-1, that is commensurate with $q=1/3$. This CDW arises via a lockin transition from an incipient, slightly incommensurate CDW with $q \sim 0.31$ \cite{Lee2019}. 

When BaNi$_{2}$As$_{2}$ is substituted with Co or Sr, the triclinic phase that hosts C-CDW-1 is suppressed and a SC dome emerges \cite{Lee2019,Eckberg2018,Eckberg2020}, suggesting C-CDW-1 plays a similar role to antiferromagnetism in chemically-substituted BaFe$_{2}$As$_{2}$. Further, full suppression of the CDW by Sr substitution, which occurs at the critical composition $x_\text{c}=0.7$, results in a quantum phase transition (QPT) characterized by nematic fluctuations that drive a sixfold enhancement of the superconducting $T_\text{c}$ \cite{Eckberg2020,Lederer2020}. Substituted BaNi$_{2}$As$_{2}$ is therefore an exciting new system in which the interaction between SC, CDW, and nematic order can be studied in detail.

Recent studies revealed several peculiar properties of Ba$_{1-x}$Sr$_{x}$Ni$_{2}$As$_{2}$ that are not fully understood \cite{Lee2019,Eckberg2020}. First, for the composition range of $0.4 < x < x_\text{c}$, the SC transition in transport measurements is broadened and occurs at a higher temperature than the specific heat anomaly \cite{Eckberg2020}. Why the resistive and thermodynamic transitions should be split in this manner is not known. Second, the IC-CDW was identified in Ref. \cite{Eckberg2020} as a lattice-driven instability, without electronic character, since it shows no precursor response in the nematic susceptibility. However, it was also observed that the elastoresistance becomes hysteretic whenever the IC-CDW is present, suggesting the two phases are coupled. These two observations are seemingly contradictory, since a purely structural phase transition should not influence the nematic response in this way. The nature of the interaction between the IC-CDW and the nematic phase remains unclear.

Here, using x-ray scattering, we present the discovery of a {\it third} CDW in Ba$_{1-x}$Sr$_{x}$Ni$_{2}$As$_{2}$, which we denote C-CDW-2, in the composition range $0.4 < x < x_\text{c}$, where $x_\text{c}=0.7$. This CDW is commensurate with a period of two lattice parameters (period-2). A CDW  with a generic wave vector has a complex order parameter \cite{mcmillan-1975}. The period-2 CDW state is special in that its ordering wave vector lies at the edge of the Brillouin zone (BZ) and, hence, its order parameter is real. Thus, the topological defects of a period-2 CDW are domain walls where the order parameter changes sign and therefore it must vanish at the location of the domain wall. We will show below that competing (weaker) SC state can be nucleated on these domain walls. The implication is that the phase at $0.4 < x < x_\text{c}$ may be a heterogeneous state in which SC resides at the topological defects of the CDW order.

Further, we report a wider x-ray momentum survey showing that the IC-CDW phase is in fact bidirectional with 90$^\circ$ rotational symmetry. We observed additional satellite reflections demonstrating a coherent ``4$Q$" state with $C_4$ symmetry that does not break the rotational symmetry of the underlying tetragonal lattice. This observation explains the absence of a precursor nematic response in elastoresistivity measurements \cite{Eckberg2020}, and suggests this CDW could be electronic and strongly coupled to the nematic order \cite{Eckberg2020}.

Single crystals of Ba$_{1-x}$Sr$_{x}$Ni$_{2}$As$_{2}$ with $x=0$, 0.27, 0.42, 0.47, 0.65, and 0.73 were measured in this study. Three-dimensional x-ray surveys of momentum space were obtained using a Mo $K_{\alpha}$ (17.4 keV) microspot x-ray source and a Mar345 image plate detector by sweeping crystals through an angular range of 20$^\circ$. All temperature evolution measurements in this study are conducted while warming the samples starting from low temperature (see Supplemental Material for experimental details \cite{supplement}).

\begin{figure}
\includegraphics{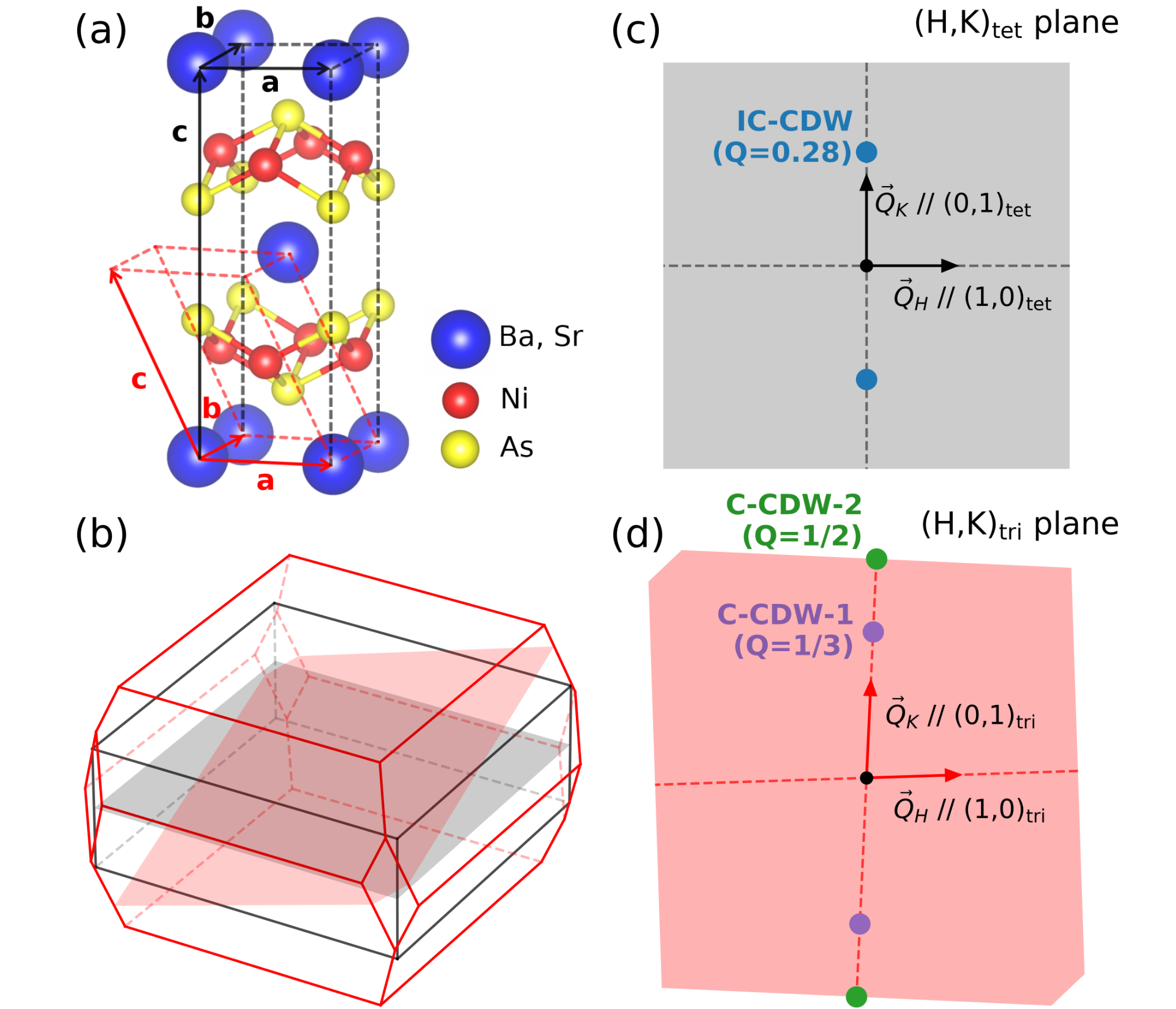}
\caption{\label{fig:fig1} Tetragonal and triclinic structure of Ba$_{1-x}$Sr$_{x}$Ni$_{2}$As$_{2}$. (a) The crystal structure and the unit cells of tetragonal (black dashed lines) and triclinic phases (red dashed lines). (b) The BZ boundaries of tetragonal (black lines) and triclinic (red lines) phases. The planes colored in black and red represent $H$-$K$ planes in the tetragonal and triclinic phases, respectively. (c), (d) The $H$-$K$ planes of tetragonal and triclinic phases, respectively, showing the location of three CDWs in momentum space in each phase. 
}
\end{figure}

The parent compound, BaNi$_{2}$As$_{2}$, has tetragonal $I4/mmm$ symmetry at room temperature and undergoes a phase transition to triclinic $P\bar{1}$ at $T_{\text{s}}=136$ K \cite{Sefat2009, Lee2019}. This transition is observed in x-ray measurements as a splitting of the tetragonal reflections into four peaks due to the formation of triclinic twin domains \cite{Sefat2009,Lee2019,supplement}. Measurements in the triclinic phase were indexed by selecting a single subset of these four reflections, emphasizing scattering from a single domain. Figures \ref{fig:fig1}(a) and (b) show the conventional unit cells and the BZ boundaries in the tetragonal and triclinic phases. Note that the $H$-$K$ planes of these two phases are not parallel, but are tilted by about 20$^\circ$ from one another. Throughout this letter, we use $(H, K, L)_{\mathrm{tet}}$ and $(H, K, L)_{\mathrm{tri}}$ to denote momentum space locations indexed with the tetragonal and triclinic unit cells, respectively (see Supplemental Material for more details on the structure and indexing \cite{supplement}). 

Here, we find that when BaNi$_{2}$As$_{2}$ is substituted with Sr, the triclinic transition as seen with x-rays initially increases in temperature, reaching a maximum $T_\text{s} = 141$ K at $x=0.27$. Further substitution decreases $T_\text{s}$ until, at $x=0.73$, the structure transition is no longer observed, indicating a quantum phase transition at $x_\text{c} \sim 0.7$, consistent with conclusions from transport experiments \cite{Eckberg2020}. In Fig. \ref{fig:fig5} we compare our results for the triclinic transition to the transport phase diagram of Ba$_{1-x}$Sr$_{x}$Ni$_{2}$As$_{2}$. The triclinic transition determined by x-ray scattering while warming (purple line) is slightly higher in temperature than the cooling transition determined by transport (black dashed line) \cite{Eckberg2020}, demonstrating the first order nature of the transition. Nevertheless, we see no evidence for coexistence of tetragonal and triclinic phases, unlike Co-substituted compounds, Ba(Ni$_{1-x}$Co$_{x}$)$_{2}$As$_{2}$, in which an extended region of coexistence occurs \cite{Lee2019}. 

\begin{figure*}
\includegraphics{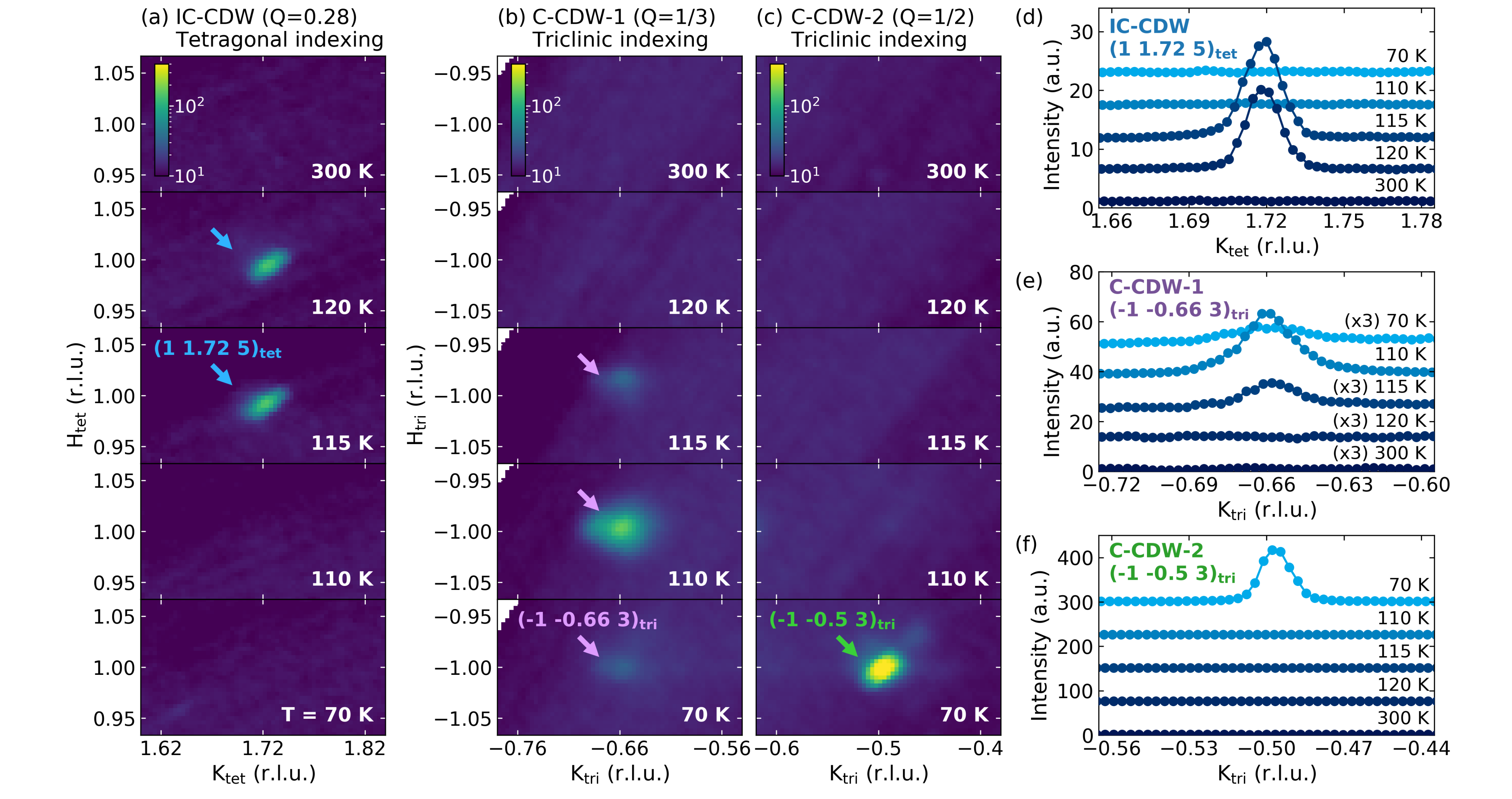}
\caption{\label{fig:fig2} Three distinct CDWs in Ba$_{0.58}$Sr$_{0.42}$Ni$_{2}$As$_{2}$. (a), (b), (c) $H$-$K$ maps of IC-CDW, C-CDW-1, and C-CDW-2, respectively, at a selection of temperatures. (d), (e), (f) Line momentum scans of the CDW reflections shown in (a), (b), and (c), respectively, along the corresponding modulation direction.
}
\end{figure*}

The primary result of this study is the discovery in the composition range $0.4 < x< x_c$ of a third CDW, which we denote C-CDW-2. This CDW is distinct from the IC-CDW and C-CDW-1 phases reported previously \cite{Lee2019, Eckberg2020}. Figure \ref{fig:fig2} shows x-ray measurements of the sample with $x=0.42$, in which all three CDWs are observed sequentially upon cooling. The IC-CDW forms at the highest temperature, at $T_\mathrm{IC} = 132.5 \pm 2.5$ K. Its wave vector, shown in the (1,1,5) BZ in Figs. \ref{fig:fig2}(a),(d), is $q=(0, 0.28, 0)_{\mathrm{tet}}$, which is the same reported in the parent compound \cite{Lee2019}. No corresponding reflection was observed at $(0.28, 0, 0)_{\mathrm{tet}}$ in this zone, which previously led us to the conclusion that the IC-CDW is unidirectional \cite{Lee2019}. Below, we present data revising this conclusion. 

Upon further cooling, the IC-CDW is replaced by C-CDW-1 at $T_{\text{C1}} = 117.5 \pm 2.5$ K (Figs. \ref{fig:fig2}(b),(e)), which coincides with the triclinic transition. Its wave vector is $q=(0, 1/3, 0)_{\mathrm{tri}}$, which is commensurate with a period of three lattice parameters. As in the case of IC-CDW, no $(1/3, 0, 0)_{\mathrm{tri}}$ peak was observed. This CDW is also observed in Co-substituted compounds, where it exhibits a lock-in effect \cite{Lee2019}. However, we observe no lock-in effect in Sr-substituted materials.
 
A third, previously unobserved CDW, which we call C-CDW-2, appears at lower temperature, $T_{\text{C2}} = 95 \pm 5$K (Figs. \ref{fig:fig2}(c),(f)). Its wave vector, $q=(0, 1/2, 0)_{\mathrm{tri}}$ is commensurate with a period of two lattice parameters. Again, no corresponding peak was observed at $q=(1/2, 0, 0)_{\mathrm{tri}}$. At $T_{\text{C2}}$, the intensity of C-CDW-1 drastically drops, and the intensity C-CDW-2 continuously increases down to our base temperature. We summarize the momentum space locations of all three CDWs in Fig. \ref{fig:fig1}(b)-(d). We emphasize that these three CDWs reside in very different locations in momentum space; the difference between the three CDW wave vectors is not merely due to a change of indexing coordinates. 

We now discuss the rotational symmetry of the three CDWs. The satellite reflections in all three phases appear in only one direction in a given BZ, which would normally be interpreted as evidence that all three CDWs are unidirectional. This is unsurprising for C-CDW-1 and C-CDW-2 since they reside in the triclinic phase in which $C_4$ (rotational) symmetry is broken by the underlying lattice. However, it is puzzling that IC-CDW should also be unidirectional, since it resides in the tetragonal phase where $C_4$ symmetry is preserved. Our claim that IC-CDW is unidirectional \cite{Lee2019} led the authors of Ref. \cite{Eckberg2020} to conclude that it is a purely structural phase transition, uncoupled to the valence electron system, since they observed no precursor response in the nematic susceptibility expected of an electronic phase with broken rotational symmetry. 

\begin{figure}
\includegraphics{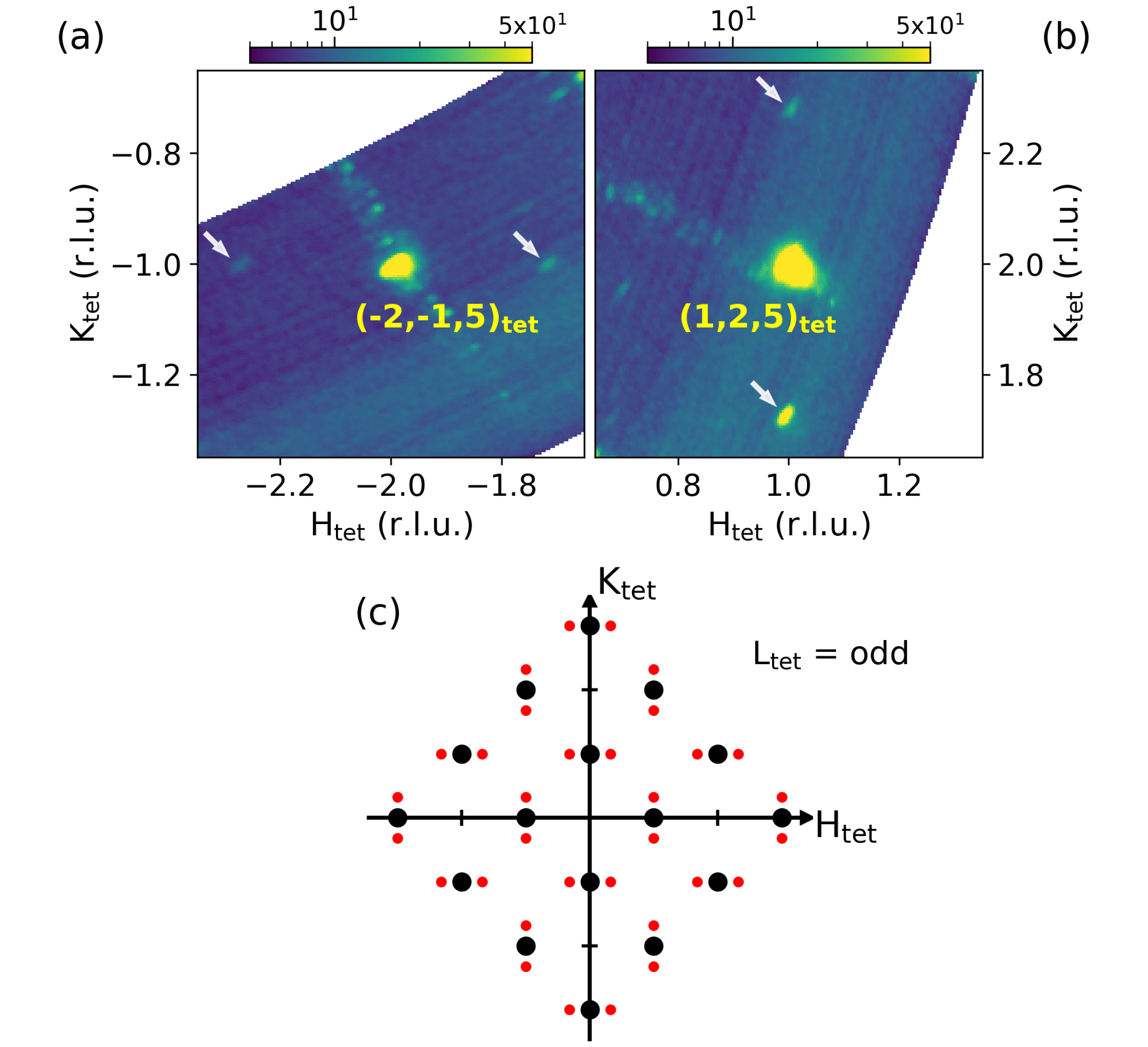}
\caption{\label{fig:fig3} Bidirectionality of the IC-CDW phase. (a) $H$-$K$ map around $(-2,-1,5)_\text{tet}$ showing the IC-CDW satellites at $(\pm 0.28, 0, 0)_\text{tet}$. The satellites are absent at $(0, \pm 0.28, 0)_\text{tet}$. (b) $H$-$K$ map around $(1,2,5)_\text{tet}$ showing the satellites at $(0, \pm 0.28, 0)_\text{tet}$. The satellites are absent at $(\pm 0.28, 0, 0)_\text{tet}$. (c) Illustration of the symmetry pattern of IC-CDW in $H$-$K$ plane at odd-numbered $L$. Black dots represent Bragg peak locations, and red dots represent IC-CDW peak locations. 
}
\end{figure}

We reexamined this issue by performing a wide, three-dimensional x-ray survey of a 20$^\circ$ wedge of momentum space, analyzing where CDW satellites reside in multiple BZs. We found that the C-CDW-1 and C-CDW-2 show the same unidirectionality in all zones observed (see Supplemental Material \cite{supplement}), affirming that these triclinic CDWs are unidirectional as claimed \cite{Lee2019}. However, the IC-CDW exhibits the more complicated pattern shown in Fig. \ref{fig:fig3}. While any given BZ has only two satellites, their orientation is different in different BZs. The overall pattern is invariant under 90$^\circ$ rotations around the origin, and therefore exhibits the same $C_4$ symmetry as the underlying tetragonal lattice. We conclude that the IC-CDW is not unidirectional, but is a coherent ``4$Q$" state in which two reflections in each BZ are extinguished by some additional symmetry. This result implies that IC-CDW may be electronic in origin after all, and may couple strongly to the nematic order. 

\begin{figure}
\includegraphics{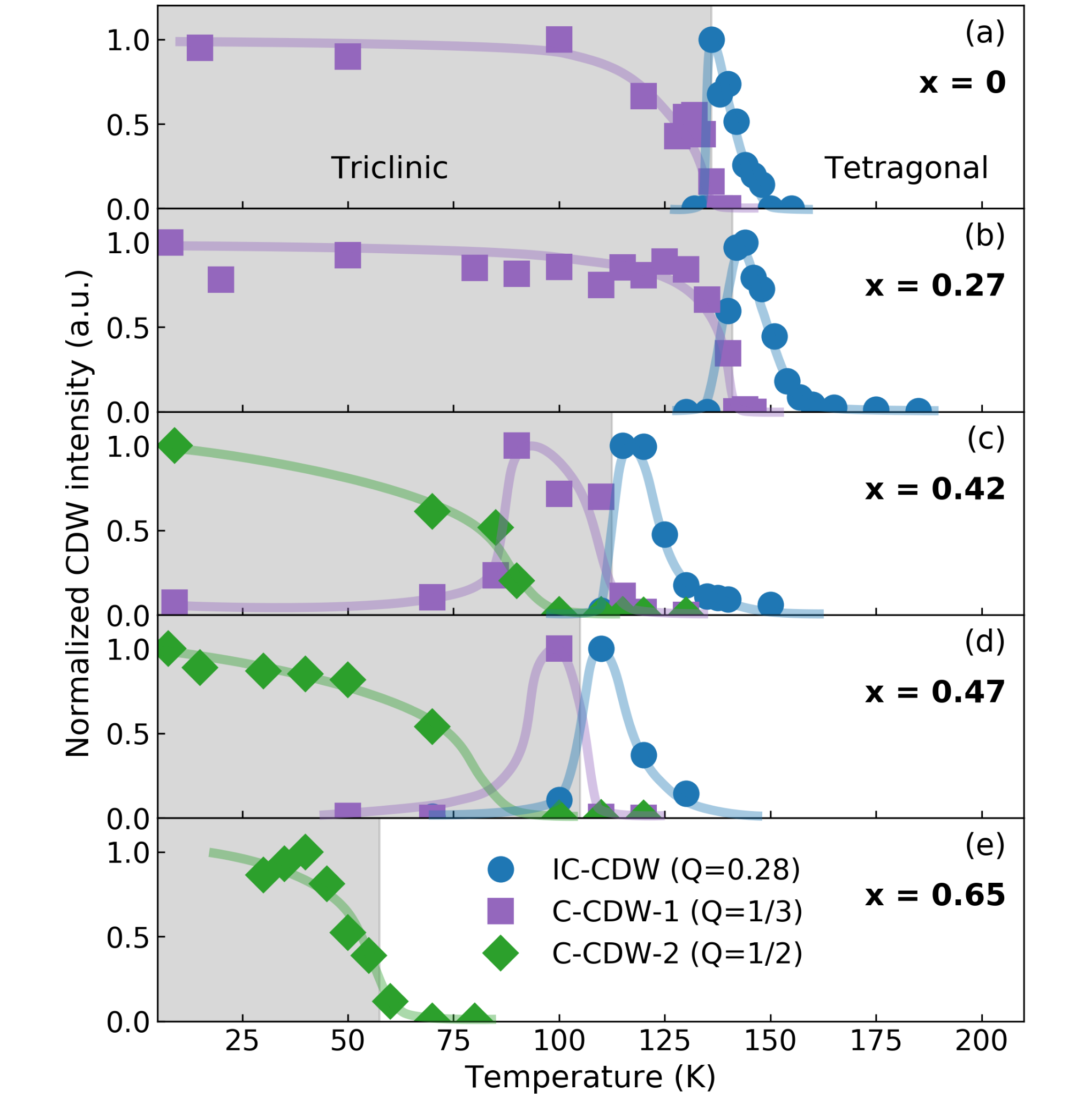}
\caption{\label{fig:fig4} Warming temperature dependence of the integrated intensities of the IC-CDW (blue circles), C-CDW-1 (purple squares), and C-CDW-2 (green diamonds) reflections in Ba$_{1-x}$Sr$_{x}$Ni$_{2}$As$_{2}$. The Sr composition, $x$, is shown in each panel. The shaded region represents the temperature range of the triclinic phase. Each curve is scaled to the maximum CDW intensity at low temperature. The solid lines are guides to the eye.
}
\end{figure}

Figure \ref{fig:fig4} summarizes the behavior of the three CDWs over the full range of Sr composition investigated. The IC-CDW is present from $x=0$ to $0.47$ and is strongest within a $\sim$20 K range above the tetragonal-triclinic phase boundary. The C-CDW-2 phase is first observed at $x=0.42$, where it replaces C-CDW-1 in the triclinic phase, and persists up to $x=0.65$ \cite{collini}. At $x=0.73$, no CDW transition is observed. No two CDWs are ever optimized at the same composition or temperature, indicating that the three phases likely compete with one another in the Landau sense. 

\begin{figure}
\includegraphics{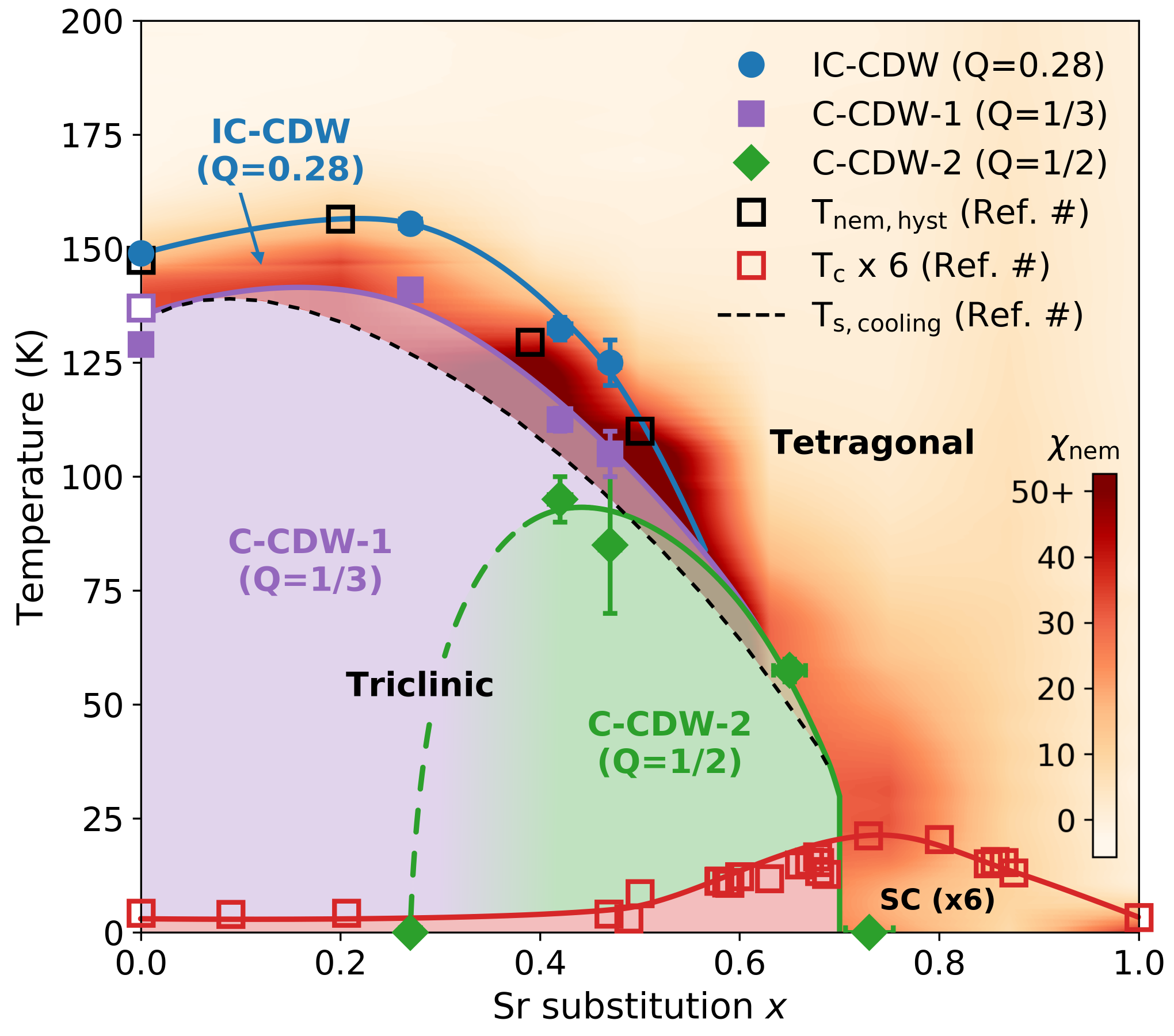}
\caption{\label{fig:fig5} Phase diagram of Ba$_{1-x}$Sr$_{x}$Ni$_{2}$As$_{2}$ comparing the IC-CDW (blue circles), C-CDW-1 (purple squares), C-CDW-2 (green diamonds) to the transport measurements of Ref. \cite{Eckberg2020}. The hollow purple square at $x=0$ represents the onset of the incipient CDW that undergoes a lock-in transition and evolves to C-CDW-1. The triclinic phase boundary on warming coincides with the boundary between C-CDW-1 and C-CDW-2 and the tetragonal phase. The cooling transition, $T_{\mathrm{s, cooling}}$ (dashed black line), determined by transport measurements shows the hysteresis region (shaded). The overlaid color scale represents the nematic susceptibility, $\chi_{\mathrm{nem}}$, and the onset temperature of the strain-hysteresis of the nematic response, $T_\text{nem,hyst}$, (black hollow squares) are plotted together. The superconducting transition temperatures, $T_{\mathrm{c}}$ (red hollow squares), determined by transport measurements are shown for comparison. $\chi_{\mathrm{nem}}$, $T_\text{nem,hyst}$, $T_{\mathrm{s, cooling}}$, and $T_{\mathrm{c}}$ are from Ref. \cite{Eckberg2020}.
}
\end{figure}

A summary phase diagram of Ba$_{1-x}$Sr$_{x}$Ni$_{2}$As$_{2}$ is presented in Fig. \ref{fig:fig5}. The anomalous superconducting phase at $0.4 < x < x_c$, in which transport and thermodynamic signatures occur at different temperatures \cite{Eckberg2020}, resides entirely within the C-CDW-2 phase. By contrast, at $x > x_c$ when no CDW is present, the superconducting signatures are normal. This implies that the peculiar superconducting state reported in Ref. \cite{Eckberg2020} is connected to the presence of C-CDW-2. 

A period-2 CDW competing with superconductivity is prone to developing a heterogeneous mixed state (see Supplemental Material Sec. V \cite{supplement}). A CDW is highly sensitive to disorder \cite{Imry-Ma-1975}, trace amounts of which can lead to the formation of domain walls with a $\pi$ phase shift, across which the order parameter changes its sign and, thus, vanishes. Since the  CDW is suppressed at the domain walls, these locations are favorable for competing superconductivity to emerge, resulting in a heterogeneous or filamentary superconducting state. In turn, Josephson coupling between these superconducting regions then leads to a globally coherent state at lower temperatures. Thus, the onset of the resistive transition is where the CDW domain walls become superconducting, with macroscopic superconductivity being achieved at lower temperatures, driven by the Josephson coupling between the domain walls. A similar mechanism was previously proposed in TiSe$_2$ \cite{Joe-2014}, which also exhibits a period-2 CDW (see also Ref. \cite{Chen-2019}), as well as in cuprate superconductors in high magnetic fields \cite{Yu-2019}. This phenomenon bears a close analogy with the CDW order seen in the SC halos of vortices of high $T_c$ superconductors  \cite{Hoffman-2002,Kivelson-2002}, and iron-based superconductors with structural twin domains \cite{Kalisky2010} or antiferromagnetic domains \cite{Xiao2012}. We therefore identify this unusual SC state as a heterogeneous state in which SC is locally formed at domain walls of C-CDW-2 that are consequence of any disorder present in the system.

Another feature of the phase diagram (Fig. \ref{fig:fig5}) is that there is a close correspondence between the presence of the IC-CDW and irreversible behavior in the nematic susceptibility measured with elastoresistance techniques \cite{Eckberg2020}. When the IC-CDW is present, the strain response is hysteretic. When absent, the behavior is reversible. As discussed above, it is possible that the IC-CDW couples strongly to the nematic order parameter, and therefore pins the nematic fluctuations. The hysteretic response then can be understood as the training of these static nematic domains by the applied elastic strain field. When the IC-CDW is absent, the nematic domains become dynamic, the response becomes reversible, and the fluctuations enhance superconductivity near $x_c=0.7$, as claimed in Ref. \cite{Eckberg2020}. 

In summary, we showed that Ba$_{1-x}$Sr$_{x}$Ni$_{2}$As$_{2}$ exhibits three distinct charge density waves, IC-CDW, C-CDW-1 and C-CDW-2. The order parameter of C-CDW-2, which is period-2, is always zero at antiphase domain walls, allowing the competing superconductivity to emerge locally. This results in a heterogeneous superconducting state for $0.4 \leq x \leq x_{\mathrm{c}}=0.7$. We also showed that the IC-CDW can strongly couple to the nematic order parameter, and promote static nematic domains by pinning the fluctuations. Our study establishes BSNA as a rare example of a material containing three distinct CDWs, and an exciting testbed for studying coupling between CDW, nematic, and SC orders.

\begin{acknowledgements}
We thank P. Ghaemi for an early collaboration, and S. A. Kivelson for discussions.
X-ray experiments were supported by the U.S. Department of Energy, Office of Basic Energy Sciences grant no. DE-FG02-06ER46285 (PA). Theory work was supported in part by the U.S. National Science Foundation grant DMR 1725401 (EF). Materials synthesis was supported by the National Science Foundation Grant no. DMR1905891 (JP). P. A. and J. P. acknowledge the Gordon and Betty Moore Foundation’s EPiQS Initiative through grant nos. GBMF9452 and GBMF9071, respectively. 

\end{acknowledgements}

\bibliography{BSNAbib}

\end{document}